\documentclass[%
 reprint,
 amsmath,amssymb,
 aps,
 intlimits,
pra,
]{revtex4-2}

\usepackage{import}
\usepackage{graphicx}
\usepackage{dcolumn}
\usepackage{bm}
\usepackage[colorlinks=true,linkcolor=red]{hyperref}%

\def\beq{\begin{equation}}
\def\eeq{\end{equation}}
\def\bea{\begin{eqnarray}}
\def\eea{\end{eqnarray}}
\def\brcl{\begin{array}{rcl}}
\def\bccl{\begin{array}{ccl}}
\def\blcl{\begin{array}{lcl}}
\def\err{\end{array}}

\def\fatr{{\bf r}}

\begin{document}

\title{
Simplifying inverse material design problems for fixed lattices with alchemical chirality
}

\author{Guido Falk von Rudorff}
\affiliation{Faculty of Physics, University of Vienna, 1090 Vienna, Austria}
\affiliation{Institute of Physical Chemistry and National Center for Computational Design and Discovery of Novel Materials (MARVEL), Department of Chemistry, University of Basel, 4056 Basel, Switzerland}

\author{O. Anatole von Lilienfeld}
\email{anatole.vonlilienfeld@univie.ac.at}
\affiliation{Faculty of Physics, University of Vienna, 1090 Vienna, Austria}
\affiliation{Institute of Physical Chemistry and National Center for Computational Design and Discovery of Novel Materials (MARVEL), Department of Chemistry, University of Basel, 4056 Basel, Switzerland}

\date{\today}

\begin{abstract} 
Massive brute-force compute campaigns relying on demanding {\em ab initio} calculations routinely
 search for novel materials in chemical compound space, 
the vast virtual set of all conceivable stable combinations of elements and structural configurations which form matter. 
Here we demonstrate that 4-dimensional chirality, arising from anti-symmetry of alchemical perturbations, dissects that space
and defines approximate ranks which effectively reduce its formal dimensionality, and enable us to 
break down its combinatorial scaling. 
The resulting distinct `alchemical' enantiomers 
must share the exact same electronic energy up to third order  --- independent of respective covalent bond topology,
and imposing relevant constraints on chemical bonding.  
Alchemical chirality deepens our understanding of chemical compound space and enables the `on-the-fly' establishment of new trends without empiricism
for any materials with fixed lattices. 
We demonstrate its efficacy for three such cases: 
i) new formulas for estimating electronic energy contributions to chemical bonding;
ii) analysis of the perturbed electron density of BN doped benzene;
and iii) ranking stability estimates for BN doping in over 2,000 naphthalene and over 400 million picene derivatives.
\end{abstract}

\maketitle



\section{Introduction}
\begin{figure*}
    \centering
\includegraphics[width=\textwidth]{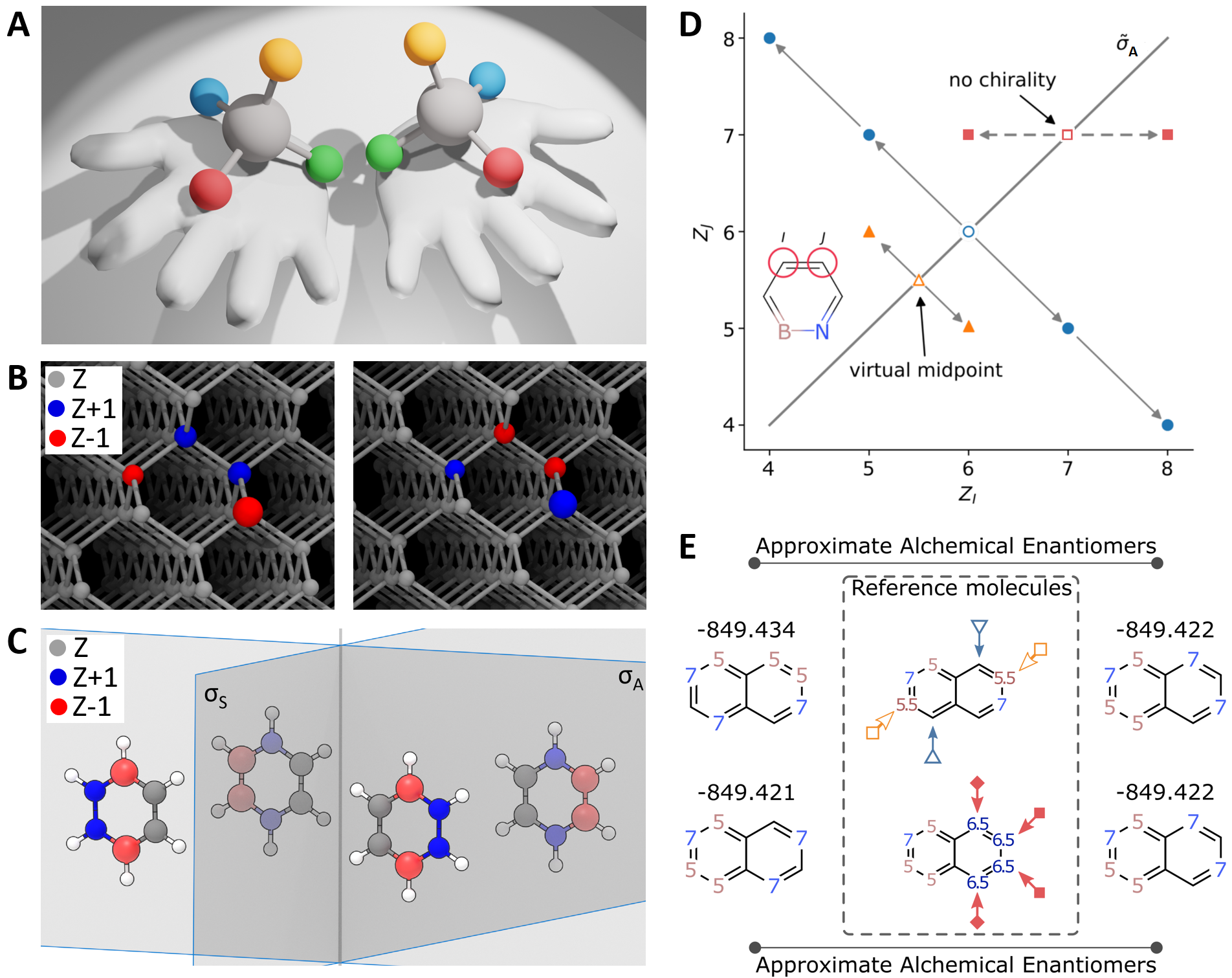}
    \caption[]{A: Enantioselective catalysis enables the synthesis of either conventional enantiomer, related to its counterpart through spatial reflection symmetry. 
    B: Alchemical (not spatial) chirality at four sites in a diamond cubic lattice relates BN doped alchemical enantiomers through alchemical reflection (Boron, nitrogen, and carbon in red, blue, and gray, respectively with their nuclear charge differences). 
    C: Alchemical reflection plane $\sigma_A$ for the six carbon nuclei ($Z$ = 6) in benzene connecting 1,4,2,3-tetrahydrodiazadiborine (THDADB) to 1,2,3,6-THDADB compared to the spatial reflection plane $\sigma_S$. Boron, nitrogen, carbon and hydrogen in red, blue, gray, and white, respectively.
    D: Illustration of approximate alchemical reflection plane $\tilde\sigma_A$ (diagonal) in the space defined by the nuclear charges $Z$ of two atoms $I,J$ in three cyclic CBNC-containing reference molecules (empty symbols). 
    Three reflections via virtual reference with fractional nuclear charges (orange triangle), real reference with two carbons (blue circle), and real reference with two nitrogens (red square) are indicated. 
    The latter reflection does not yield alchemical enantiomers.
    E: Chains of approximate alchemical enantiomers. Top left and top right molecules are connected via one reference molecule while bottom left and bottom right molecule are connected via another reference molecule. Note that both molecules on the right are identical, and therefore the three distinct molecules form a chain of approximate alchemical enantiomers. 
    Transmutating atoms and symmetry-equivalent sites are denoted with nuclear charges and identical arrows, respectively. Electronic energies given in Hartrees.
    }
    \label{fig:schematicrules}
\end{figure*}

The computationally demanding virtual simulation of molecules and materials, performed to predict their physical, materials and chemical properties, has become a routine tool in the molecular and materials sciences. Current efforts geared towards computational materials and molecular design might enable one day the realization of the holy grail of automatized experimental design and discovery. 
Driven by the accelerating progress of compute hardware and statistical learning (artificial intelligence), first seminal examples of integrating sophisticated software and robotics to perform experimental sequences and to establish rules and trends among properties and materials, as well as their synthesis, {\em in realiter} have recently been introduced~\cite{ShrierNorquist2016ML4failedExperiments,Cronin2020AutonomousRobot,Cronin2020RoboticTrends,Aspuru2020AutonomousExperimentation,Aspuru2020selfdrivingDiscoveryThinFilms}. 
However, the lofty goal of `materials on demand' has still remained elusive, even when doing it just {\em in silico}.

The use of empirical trends to guide experimental design has had a long tradition in the chemical sciences. Popular examples
include Mendeleev's discovery of the periodic table, Hammett's relationship, Pettifor's numbering scheme,
Bell-Evans-Polanyi principle, Hammond's postulate, or Pauling's covalent bond postulate~\cite{pauling1932CovalentBondsPNAS}. 
Modern systematic attempts to establish and exploit such rules in terms of quantitative
structure-property relationships have led to computationally advanced bio-, chem-, and materials-informatics methodologies~\cite{AspuruOpreaRoitbergTetko2020qsar}.
Unfortunately, these methods are typically inherently limited to certain applicability domains, 
and do not scale due to their empirical nature~\cite{SchneiderReview2010}. 
To rigorously explore the high-dimensional chemical compound space (CCS)~\cite{mullard2017drug}, i.e.~the combinatorially scaling number of all conceivable molecules or materials (usually
defined by composition, constitution, and conformation), the quantum mechanics of electrons 
ought to be invoked. 

It is thus not surprising that {\em ab initio} based materials design approaches have been at the forefront for more than 20 years~\cite{Beratan1991,ceder1998predicting,ZungerPRL1998,ZungerNature1999,alberi20182019}, and have played a major role in popularizing the use of efficient and accurate quantum methods, such as density functional theory~\cite{HK,PBE,ComputationalMaterialsDesign_MRS2006,BurkePerspectives_2012jcp,marzari2016materials}. 
Sampling CCS from scratch, even when done within efficient optimization algorithms, 
is typically an encyclopedical endeavour by nature, 
and ignores many of the  
underlying relations among different properties and materials. 
Quantum machine learning models~\cite{QMLessayAnatole} statistically exploit such implicit correlations, 
hidden in the data, and have been successfully used to accelerate
CCS exploration campaigns~\cite{Anatole2020NatureReview}.
Machine learning efficiency and transferability demonstrably benefits
greatly from explicitly enforcing known relationships 
(e.g.~translational, rotational, or atom index invariances) directly 
in the model construction, rather than having to learn them
agnostically from data~\cite{MachineLearningMeetsQuantumPhysics2020book}. 
Specific examples include explicitly imposing forces and curvatures in the loss function~\cite{RampiMLQMMM,CovariantKernelsSandro2016,chmiela2017machine},
spatial symmetry relations~\cite{grisafi2018symmetry,chmiela2019sgdml},
or arbitrary differential relations~\cite{OQML}.
But even for the most efficient and transferable statistical models, e.g.~the atom-in-molecule
fragment based approach~\cite{Huang_2020},
the acquisition of training data in sufficient quantity and quality 
requires considerable up-front investments.

In this paper, we introduce the fundamental notion of a new symmetry relation in CCS which is fully consistent with the {\em ab initio} view of matter~\cite{anatole-ijqc2013},
and effectively enables us to solve the inverse materials design problem in a non-empirical and highly efficient manner. 
Spatial symmetry considerations have been crucial for the unravelling of 
some of the most fundamental laws of nature and are heavily used in many fields. 
In {\em ab initio} calculations, for example, symmetry group 
theory arguments are common to reduce computational complexity and load. 
Symmetry constraints on compositional degrees of freedom would be highly desirable in order to establish general rules among distinct materials and properties, and to generally improve
our understanding of chemical compound space~\cite{anatole-ijqc2013}. 

In analogy to conventional spatial chirality, we here define `alchemical chirality' as a reflection plane in the space spanned by nuclear charges at fixed atomic positions as they enter the electronic Schr\"odinger equation.
An illustrative comparison is given in Fig.~\ref{fig:schematicrules} (panels A and B), for conventional enantiomers consisting of a tetra-valent carbon atom with four different substitutions and for alchemical enantiomers consisting of doubly BN doped carbon in the diamond crystal structure. Fig.~\ref{fig:schematicrules}C compares and relates this newly described alchemical reflection $\sigma_A$ with the conventional spatial reflection $\sigma_S$ for the same dopant pattern as in Fig.~\ref{fig:schematicrules}B for a single molecular skeleton. In this case, four subsequent reflections alternating between alchemical and spatial reflections return to the original molecule. While any spatial reflection leaves the molecule unchanged, an alchemical reflection affects the nuclear charges and therefore creates a different molecule as reflection image.

Exchange of the dopant atom sequence in Fig.~\ref{fig:schematicrules}B from NBBN $\rightarrow$ BNNB is  an alchemical reflection around pristine diamond: for each site, the nuclear charge difference to diamond gets inverted. As such, in the space spanned by all nuclear charges of the system, pristine diamond corresponds to a reflection center. All reflections that leave the total nuclear charge unchanged are defined by a hyperplane (cf. Figure 1D) which we refer to as the nuclear charge reference plane. 
In other words: Treating the change from pristine diamond to the two doped variants as a perturbation of the system Hamiltonian, there is an anti-symmetry relation between these alchemical perturbations.


No other spatial symmetry operation (rotation, reflection, or inversion) can interconvert the constitutional isomers in Fig.~\ref{fig:schematicrules}B, thereby necessitating a fourth dimension, namely the nuclear charges. This yields 4-dimensional alchemical chirality. Note that chirality crucially depends on dimensionality, e.g.~the letter L is chiral within 2 dimensions only. The chiral center of that operation in itself is a compound (diamond in Fig.~\ref{fig:schematicrules}B and benzene in  Fig.~\ref{fig:schematicrules}C).

Alchemical enantiomers exist only if distinct atoms can be mapped onto each other under a symmetry operation, a consequence of the reflection in nuclear charge space that defines alchemical enantiomers. This implies that the sum over all nuclear charges of alchemical enantiomers is identical (see Fig.~\ref{fig:schematicrules}D).
If a compound has no spatial symmetry, atom sites with similar electron density derivatives $\partial^n_Z \rho$  constitute these pairs just as strictly symmetry-equivalent atoms do. Alchemical chirality requires a one-to-one correspondence of sites with opposite change in nuclear charge. At least two such pairs need to exist in a compound to obtain alchemical enantiomers which are different constitutional isomers. For a single pair of symmetry-equivalent atoms, the alchemical reflection in nuclear charge space would trivially connect two spatially symmetric compounds. 

In summary: \textit{Exact alchemical enantiomers} are defined as two spatially non-superimposable,
alchemically coupled,
and iso-electronic compounds 
with the same formal charge,
where each transmutating atom is assigned to exactly one subset within each of which averaging of nuclear charges results in identical chemical environments. 
Note that alchemical enantiomers can differ in chemical compositions, and that any given alchemical enantiomer can have as many alchemical mirror images as different valid averaged reference compounds can be defined. 
An alchemical enantiomer and all its possible mirror images are degenerate in the electronic energy up to 3rd order.
\textit{Approximate alchemical enantiomers} 
differ from their exact analogue in that
the 
averaging results in similar,
rather than identical, chemical environments for all
transmutating atoms.
Alchemical enantiomers have approximately the same electronic energy (see Theory section below). This symmetry is then broken by (a) the nuclear-nuclear repulsion and (b) geometry relaxation into distinct total energy minima. The nuclear repulsion will typically dominate, implying that alchemical enantiomers with spatially closer atoms of higher nuclear charge will exhibit higher total energies than their respective mirror images.

While the restriction to fixed atomic positions might seem severe at first, we highlight in the following the importance and relevance of classes of fixed configurational frameworks for materials design applications. 
More specifically, we discuss the implications of alchemical chirality for system classes that are low dimensional in their structural degrees of freedom, and where the problem is dominated by the combinatorial scaling due to varying chemical composition. 
Examples abound and include any variation of graphitic motifs (studied below) prevalent in nano-technological applications, in inorganic materials such as MAX-phases, or in organic electronics, e.g.~polycyclic aromatic hydrocarbon derivatives, as well as other rigid scaffolds, such as Metal-Organic-Frameworks or perovskites. All these systems would be directly amenable to alchemical chirality-based estimates (ACE) 
i.e.~approximating the electronic energies for alchemical enantiomers as degenerate, and adding nuclear repulsions to obtain relative energies. 
This way, ACE enables the ranking and grouping of large subsets of materials.
Within any of these classes, the number of possible materials increases combinatorially with the number of building blocks. For real-world applications, defects often need to be included in the quantum chemistry model, which further increases the chemical space under consideration. For particularly rigid frameworks and extended periodic materials, only a local and low-dimensional geometric response to a change of composition is typically observed. As such, ACE amount to a complementary means to rigorously sample compositional ensembles within a given framework class. As shown below, ACE become an enabling tool for the systematic identification of structure property relations within given compound classes that further the understanding of the respective impact of the compositional degrees of freedom in materials design. Here, we exemplify this by identifying design rules from analysis of over 400 million alchemical estimates of BN-doped derivatives in the picene framework. 

\section{Results}

\subsection{Estimating Bond Energies}
The external potential energy differences between the reference compound (defining the reflection plane such as carbon atoms in diamond in Fig.~\ref{fig:schematicrules}B) and either alchemical enantiomer are exactly mirrored in magnitude (opposite sign). As such, while the corresponding electronic Hamiltonians are alchemical mirror images of each other, their respective solutions to the electronic Schr\"odinger equation
are not necessarily equal. 
More specifically, for reflections around atoms with identical chemical environments, 
we show that the electronic energy of corresponding alchemical enantiomers 
must be degenerate up to third order within alchemical perturbation density functional theory~\cite{APDFT} (See Theory section below), higher order terms carrying different signs. 
As such, approximate alchemical enantiomers are only approximately degenerate in their electronic energy.

It turns out, however, that this approximation is quite fair, and
that useful rules for chemical binding can be derived. 
For example, by enumerating all colored connected graphs that are sub-graphs of hexagonal lattices we obtain the following ACE of 
2-body interatomic bonding for the electronic energy
\bea
E_{\rm QR} & \approx & E_{\rm SR} + 0.5 \times (E_{\rm QQ}-E_{\rm SS})
\eea
where QRS correspond to three adjacent elements in the periodic table. 
As such, this rule confirms the naive expectation that given two hetero-atomic bonds QR and SR to the same reference atom R, that electronic contribution to bonding will be larger which contains the hetero-atom with the larger electronic homo-atomic bond, i.e. QQ vs.~SS, respectively. We believe that this simple rule was hidden in the past due to the obfuscation caused by the contributing nuclear repulsion term.

While seemingly reminiscent of Pauling's electronegativity based bond energy estimate~\cite{pauling1932CovalentBondsPNAS}, $E_{\rm QS} = 0.5 \times (E_{\rm QQ} + E_{\rm SS}) + \Delta \chi$, 
we stress that this ACE relation is derived from pure symmetry and
perturbation theory based considerations which, for alchemical enantiomers, are exact up to third order.
Pauling, by contrast, merely postulated his {\em Ansatz}. 
Our rule relies on bonding information 
involving three distinct and adjacent chemical elements rather than
just two, and, maybe more importantly, it pertains 
to the electronic potential energy only, i.e.~without the nuclear repulsion 
which is trivial to add {\em a posteriori}.
This rule is easily verified for the example of BC and CN bonds
in tetrahydrodiazadiborine (Fig.~\ref{fig:orders}) by simply
comparing the sum of all bond energies in either alchemical enantiomer
(resulting in $2 E_{\rm BC} + E_{\rm NN}  \approx 2 E_{\rm NC} + E_{\rm BB}$).
\begin{figure}
    \centering
    \includegraphics[width=\columnwidth]{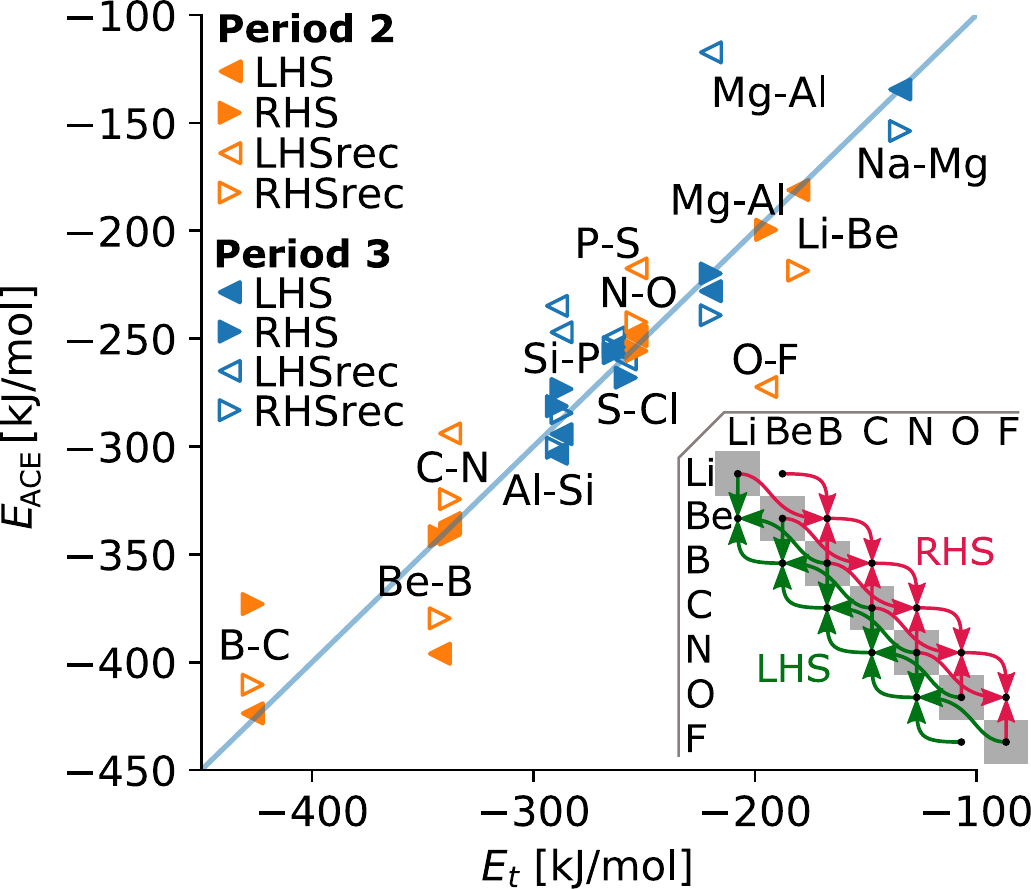}
    \caption{Predicted ($E_{\rm ACE}$) from alchemical chirality-based estimates (ACE) versus true ($E_{\rm t}$) single bond dissociation energies according to the 2-body rule for adjacent elements QRS using published DFT results for main-group elements in the 2nd and 3rd row of the periodic table 
    and assuming all bond-distance to correspond to 2 Bohr. 
    Predictions of bonding energies between QR and RS elements are denoted as LHS and RHS, respectively. Successive recursive application of the rule throughout either period is denoted by 'rec'.
    Inset: Each field in the matrix corresponds to a bond of the elements in the row and column headers. Arrows show which values when combined together yield a predicted bond energy. This shows the symmetry in going forward (RHS) and backward (LHS) through the groups in the periodic table. Using only the initial off-diagonal matrix element and then using one prediction to obtain the next following the arrows would correspond to 'rec' in the main figure.}
    \label{fig:bond}
\end{figure}

Subtracting nuclear repulsion estimates from published DFT data for all single bonds between main-group elements from the second and third period saturated by hydrogen~\cite{Elguiero2005BDEdiatomicMaingroup}, 
ACE yield bond energies of elements QR, i.e.~solely based on
bonding information of elements SR, QQ, and SS for adjacent elements QRS in the same period. 
As shown in Fig.~\ref{fig:bond}, predicting bond dissociation energies among elements R and Q or S 
(to the left or right in the period)
generates remarkably accurate predictions for bonds among elements in either period 
with a mean absolute error of $\sim$10 kJ/mol, not far from the highly coveted
`chemical accuracy' of $\sim$4 kJ/mol.
A linear fit through all the diverse chemistries encountered
yields a slope of 1.04, an off-set of $\sim$7 kJ/mol, and a 
correlation coefficient of 0.981.
Successive daisy-chaining across any given period represents a straightforward
extension which dramatically reduces the number of reference bonds required for 
ACE---but only at the cost of reduced accuracy: 
The MAE increases to $\sim$22 kJ/mol, i.e.~still  
better or on par with common generalized gradient based approximations to the 
exchange correlation potential in DFT~\cite{ChemistsGuidetoDFT}. 

Further ACE rules emerge when increasing the alchemical nuclear charge radius 
($\Delta Z = \pm1, \pm2$) with PQRST corresponding to five adjacent elements in the periodic table:
$E_{\rm PR} \approx E_{\rm TR} + (E_{\rm PP}-E_{\rm TT})/2$, 
$E_{\rm PQ} + E_{\rm QT} + 2 E_{\rm SR} \approx E_{\rm PS} + E_{\rm ST} + 2 E_{\rm QR}$,
and
$2 E_{\rm PR} + E_{\rm QT} + E_{\rm ST} \approx E_{\rm PQ} + E_{\rm PS} +2E_{\rm TR}$. 
These rules are identical for the 3D diamond lattice and the 2D hexagonal graphene structure. 
Note that other graph lattices could yield additional rules, 
and that rules for interatomic 3- and n-body contributions to binding exist as well.
For example, invoking $\Delta Z = \pm1$ only, one finds the 3-body rule that
$E_{\rm SQS} + 2 E_{\rm RSQ} + E_{\rm RQR} \approx E_{\rm QSQ} + 2 E_{\rm RQS} + E_{\rm RSR}$. 
A systematic enumeration of ACE rules for 2- and 3-body terms for graphene and diamond lattices is given in the SI.


\subsection{Ranking Molecules}
To apply ACE to inverse materials design problems, we show-case applications in well defined sub-regions of CCS and solve three specific and increasingly challenging  design tasks. All three use-cases address the combinatorial design problem of how to dope planar hexagonal lattices to an increasing extent (6, 10, and 22 carbon atoms, respectively). Hexagonal lattices are archetypical scaffolds, e.g.~relevant in the design of graphene-inspired materials for nano-technological devices~\cite{Ryczko2020}, catalytic surfaces~\cite{Keith2017alchemicalCatalysis,Keith2018benchmarkingalchemy,Keith2020acceleratedCatalystAlchemy}, porous BN-doped based nano-fibers for battery materials~\cite{PorousBNnanofibers2020graphene},
or 2D materials in general~\cite{Marzari20202highthroughput2Dmaterials}. 
Just doping with BN already results in a combinatorial explosion, already for the 77 smallest benzoid-like structures, the number of possible unique BN-doped derivatives exceeds 7 tera~\cite{Raghu2019polycyclicBNdoped}. 
As such, we believe that it is warranted, and without loss of generality, to 
focus on constitutional isomers in rigid lattices for which thermal or geometrical distortions 
can be neglected. Note that relative off-sets in total energies due to differences in stoichiometry 
are straightforward to estimate, and typically occur on different orders of magnitude, and 
that subsequent inclusion of configurational degrees of freedom within alchemical predictions is 
possible, as already demonstrated for  small molecules~\cite{Samuel-CHIMIA2014,Samuel-JCP2016} and ionic, metallic, and semi-conducting solids~\cite{MoritzBaben-JCP2016,AlchemyAlisa_2016,Samuel2018bandgaps}.
As such, the focus of the following applications lies on breaking down the
combinatorially scaling problem of energy predictions throughout colored chemical bond connectivity graphs. 


\begin{figure*}
    \centering
    \includegraphics[width=\textwidth]{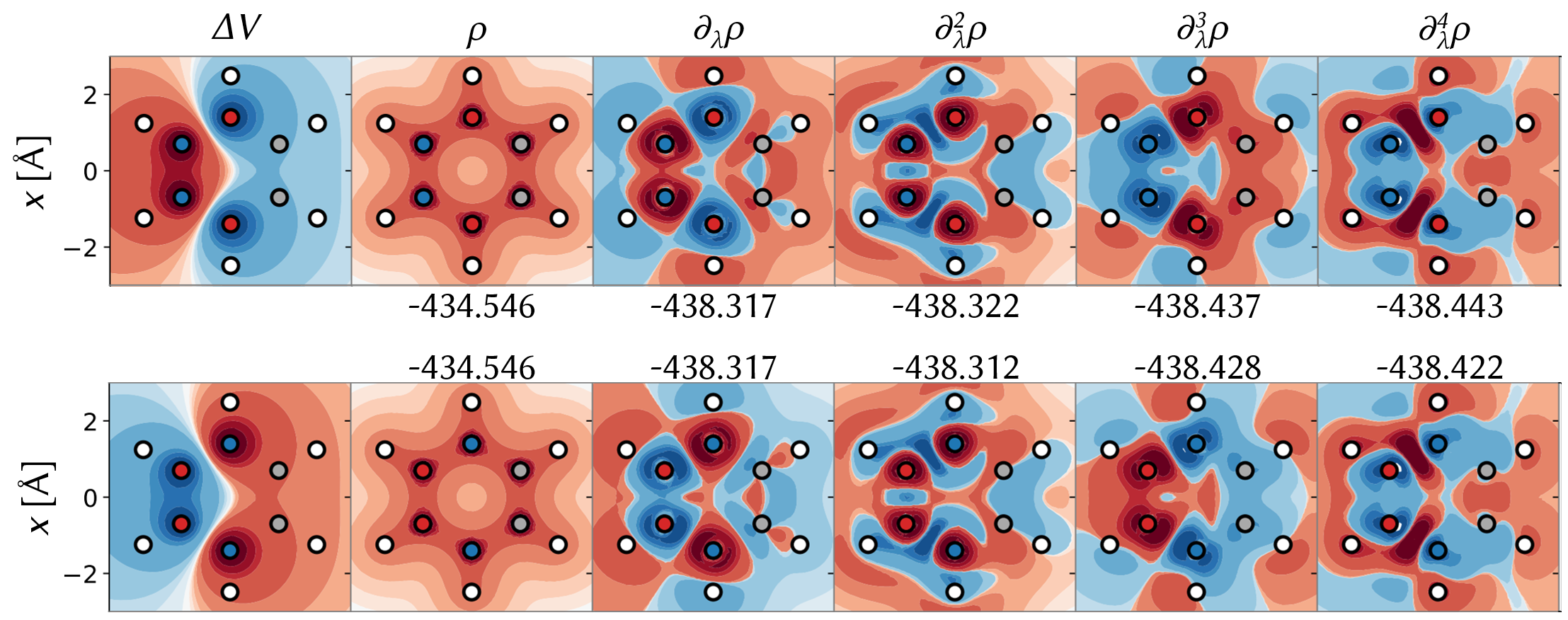}
\caption{
Alchemical enantiomers in carborazine (C$_2$H$_6$B$_2$N$_2$, tetrahydrodiazadiborine (THDADB)) with the reference molecule benzene (C$_6$H$_6$).
Left-hand column: the perturbing potential $\Delta V$ acting on benzene. 
Remaining columns, electron density derivatives $\rho$, $\partial_\lambda^{n-1}\rho$ from order zero (density of benzene) up to 4th order at CCSD/def2-TZVP level. 
Positive/negative values are shown in red/blue, respectively. All density contour lines are set at the same percentiles for each plot to render their shape comparable at different magnitudes. The percentiles are chosen to emphasize extremal values.
Next to each electron density derivative, the corresponding electronic energy estimate
by APDFT is given in Hartree for order 1 up to 5. The actual electronic energies are -438.413\,Ha (top) and -438.401\,Ha (bottom). }
    \label{fig:orders}
\end{figure*}

As a first use-case we provide an in-depth but intuitive illustration of alchemical chirality with benzene, its six equivalent carbon atoms as the mirroring reflection plane, and the alchemical enantiomers corresponding to carborazine (C$_2$H$_6$B$_2$N$_2$, tetrahydrodiazadiborine (THDADB)).
Since second order APDFT -- resulting in alchemical normal modes to define a complete basis in certain sub-spaces of CCS -- has already been applied to benzene~\cite{AlchemicalNormalModes}, the occurrence of 6 degenerate estimates among the 11 constitutional isomers of tetrahydrodiazadiborine (C$_2$H$_6$B$_2$N$_2$) can now be readily explained thanks to alchemical chirality: They represent three pairs of alchemical enantiomers. 
For the select enantiomer pair BNNBCC/NBBNCC, a molecular planar analogue to the crystalline example (Fig.~\ref{fig:schematicrules}B) is given in Fig.~\ref{fig:schematicrules}E, contrasting the conventional spatial reflection plane $\sigma_s$ with the alchemical reflection plane ${\sigma}_A$.
While NBBNCC and BNNBCC are mutually achiral in the sense of conventional spatial 3D chirality, 
the alchemical chirality relation between the two becomes obvious, with regular benzene corresponding to
the reflection plane. 



In Fig.~\ref{fig:orders}, the corresponding perturbing potentials (exact mirror images of each other) for the two benzene enantiomers from Fig.~\ref{fig:schematicrules}C are shown, as well as the resulting electron density derivatives of benzene up to 4th order. Corresponding figures for all other BN doped benzene mutants are provided in the SI. 
As it becomes obvious already by visual inspection, the perturbed densities are near identical
for even orders, and alchemically mirrored for odd orders. This indicates that the perturbational series in APDFT~\cite{APDFT} extrapolates to an approximately equal amount, leading to the near-degeneracy of the electronic energies of the two alchemical enantiomers.
To quantify this effect, Fig.~\ref{fig:orders} also shows the corresponding APDFT based electronic energy estimates up to 5th order (within perturbation theory the energy order precedes the wave-function order), numerically demonstrating that the degeneracy is identical in 2nd order, and starts to deviate by $\sim$0.01, 0.02, and 0.025 Ha for 3rd, 4th, and 5th order, respectively. The 5th order predictions deviate from the actual electronic energies by $\sim$0.03 and 0.02 Ha, respectively. 
Addition of the nuclear repulsion terms typically lifts the approximate degeneracy, 
resulting in quantitative ACE  estimates as well as simple inequality rules
for total energies that $E_{\rm BNNBCC} > E_{\rm NBBNCC}$ 
due to the larger nuclear repulsion experienced when atoms with larger nuclear charges (nitrogens in this case) are closer in proximity to each other than atoms with lower nuclear charges (boron). 

Such accuracy, while not on par with explicitly correlated quantum chemistry calculations in large basis sets, is on a similar scale as generalized gradient approximated DFT or semi-empirical quantum chemistry methods --- typically sufficiently accurate for successful computational materials design studies as demonstrated for the many examples of successful computational discovery of heterogeneous catalysts~\cite{ReviewCatalystNorskov}, or the materials project data-set~\cite{MaterialsProject}.
The following two use-cases explore this question for ACE based ranking in a more systematic and comprehensive fashion.

\begin{figure}
    \centering
    \includegraphics[width=\columnwidth]{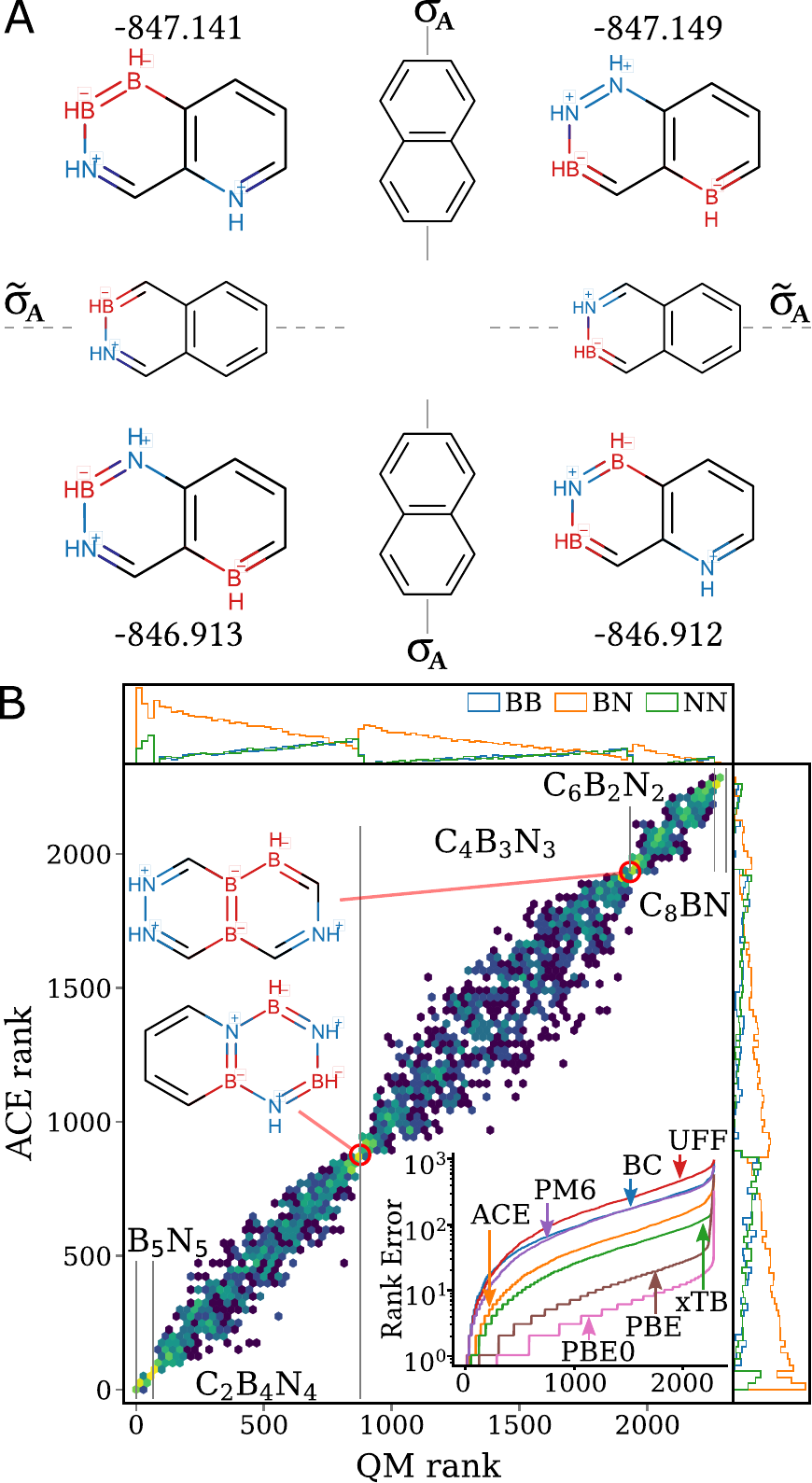}
    \caption{(A) Chains of alchemical enantiomers used for ranking molecules. Each alchemical symmetry plane is denoted with a grey line and the reference molecule, the chiral center. Exact symmetries are denoted $\sigma_A$ (stroked line) while approximate symmetries are given as $\tilde \sigma_A$ (dashed line). All CCSD electronic energies given in Hartrees.
    (B) All 2,285 unique BN-doped naphthalene derivatives ranked by their total energy as obtained from quantum chemistry calculations compared to the ranking from alchemical chirality-based estimates (ACE). Histogram colored by number of molecules in a given bin. Energy scales shown in Hartree for each stoichiometry. Energies relative to the most stable molecule. Two representative molecules of same sum formula C$_4$(BN)$_3$ and geometry are annotated. 
    Inset shows the distribution of rank errors sorted in ascending order for different methods: the force field UFF, bond counting (BC), semi-empirical PM6, ACE (this work), semi-empirical xTB-GFN2 (xTB), and the DFT methods PBE, and PBE0, both with density-fitted cc-pVDZ basis set.
    Side panels show bond type frequencies.}
    \label{fig:ranking}
\end{figure}{}
Interestingly, since the equivalence of the electronic energies is transitive, having a second reference 
that connects one alchemical enantiomer with a third one allows us to build chains of alchemical enantiomers that must have similar electronic energies, consequently enabling a ranking of molecules within one such group solely by the magnitude nuclear-nuclear interaction. 
Fig.~\ref{fig:ranking}A illustrates the chaining of exact ($\sigma_A$) and approximate ($\tilde{\sigma}_A$)
alchemical enantiomers for BN-doped naphthalene where four pairs of alchemical enantiomers could form such a chain. For  reflections along the exact alchemical symmetry axis $\sigma_A$, the energy difference is 1\,mHa or 8\,mHa, respectively. For cases with approximate alchemical symmetries $\tilde\sigma_A$, this energy difference is notably larger (237\,mHa), but typically still much smaller than the nuclear repulsion energy contributions. 

Overall, exhaustive charge-neutral and iso-electronic BN doping in naphthalene leads to 2,285 unique derivatives for which we have used eACE to establish energetic ranks. 
Having identified all alchemical enantiomers through exhaustive scanning within one stoichiometry (stoichiometries as a whole can be ranked with existing relations\cite{Mezey1998,Mezey1980,Daza2005}), we have ranked all possible molecules using ACE within groups of molecules approximately degenerate in electronic energy. 
These groups form the connected components of a graph where all molecules are nodes and only alchemical enantiomers are connected, thus allowing us to exploit the transitivity of the electronic energy degeneracy.
Within these groups, solely the nuclear-nuclear interaction induces the ranking. 
Among the groups, we rank based on averaged bond-counting results for their bond energies, where
bond-energies have been obtained by fitting to the imposed energy degeneracies among all pairs. 

Using CCSD/cc-pVDZ as a ground-truth QM method for all naphthalene derivatives on pre-optimized geometry\cite{Chakraborty2019}, we have performed an 
exhaustive validation of alchemical ranking, resulting in a Spearman correlation coefficient of 0.9899. 
Dissecting CCS in all the various possible stoichiometries of BN doped naphthalene,
the scatter plot of ACE rank vs.~QM rank results in a very reasonable correlation, 
as shown in Fig.~\ref{fig:ranking}B.
To place ACE based ranking in perspective, the inset of Fig.~\ref{fig:ranking}B also 
reports corresponding sorted ranking errors (wrt. CCSD) when using
PBE0~\cite{PBE0,PBE01,PBE02}, 
PBE~\cite{PBE}, xTB~\cite{Grimme2019GFn2xTB}, PM6\cite{PM6}, bond-counting (BC), UFF\cite{UFFRappeGoddard1992} with respective Spearman correlation coefficients of
0.9998, 
0.9983, 
0.9966, 
0.9592,
0.9562,
and
0.9021.
In terms of computational cost invested to rank all 
naphthalene derivatives, ACE ranking is slightly more expensive than bond counting, UFF or PM6
in terms of accuracy however, it clearly outperforms all three---without any
need for empirical knowledge or fitting to external data! 
It is worth noting that bond counting is the only method that has been re-parametrized on this particular data set which explains its good performance compared to PM6.
As also shown in Fig.~\ref{fig:ranking}B, ACE reproduces bond type frequency trends as a function of rank.
While the accuracy of ACE itself is not on par with more advanced semi-empirical quantum
chemistry (xTB) 
the sorted ranking error distribution suggests that 
alchemical ranking is closer to xTB than it is to PM6. 
However, in contrast to alchemical ranking, xTB relies on substantial empirical data for fitting. 
While PBE, B3LYP, and PBE0 are obviously more accurate, their computational cost 
is also three orders of magnitude larger than xTB ($\sim40$ s). 
As such, we conclude that alchemical ranking is superior to bond-counting and PM6, 
and could represent a viable alternative to xTB if accuracy thresholds are less stringent
and computational load is high. 

\begin{figure*}
    \centering
    \includegraphics[width=\textwidth]{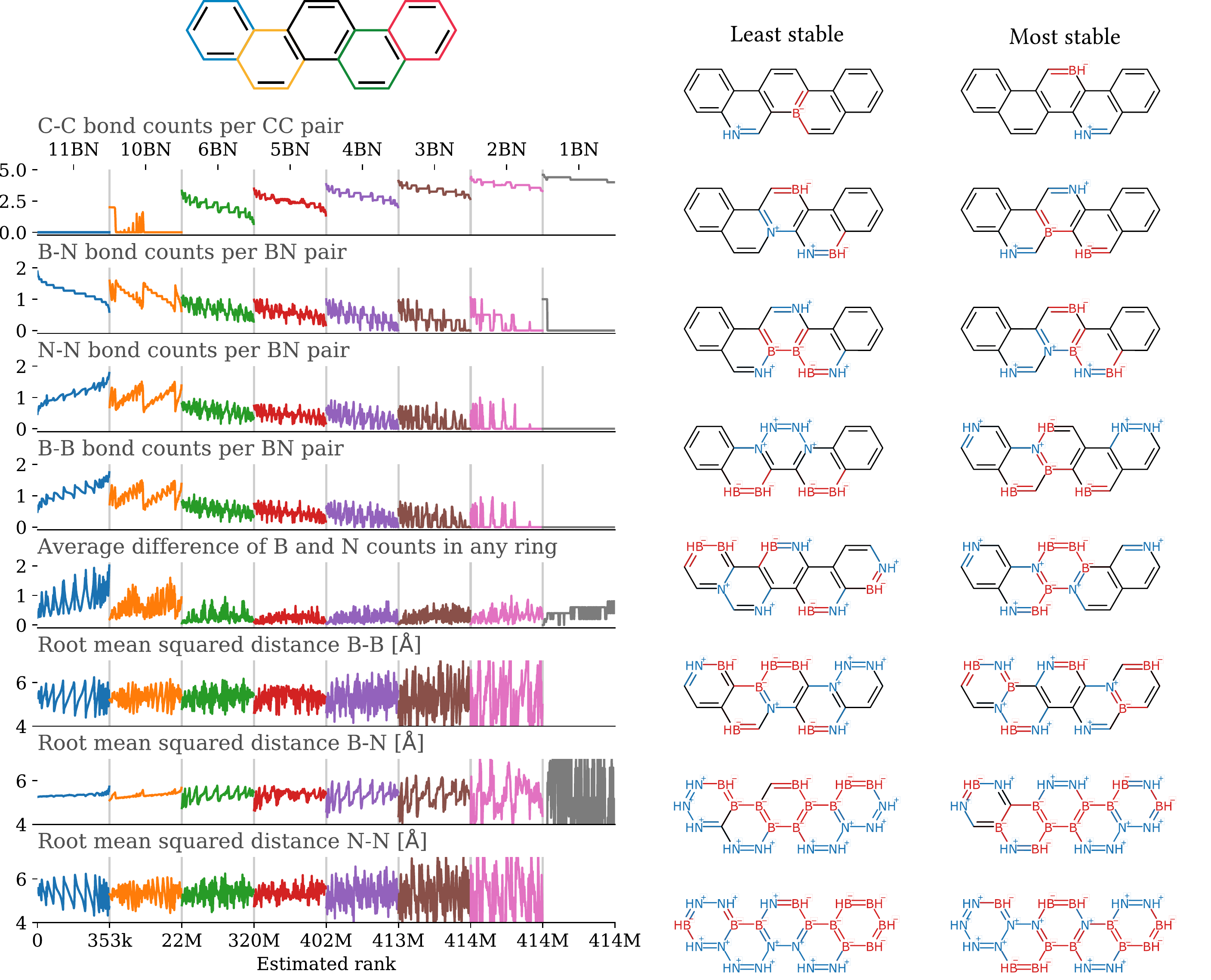}
    \caption{Trends among all 414 million BN-doped picenes of the select stoichiometries (top) ranked by their total energy based on ACE. Each row shows a geometric property for each stoichiometry as a function of the ACE rank and averaged over 200 bins. Those molecules that are the most or least stable for each stoichiometry are listed separately.}
    \label{fig:patterns}
\end{figure*}{}
As a third usecase we have used ACE to explore and deepen our understanding of the CCS spanned by the 413,887,189 unique k-fold BN doped derivatives of picene (See Fig.~\ref{fig:patterns}, with $k<7$ and $k>9$). 
Dissecting its CCS first by all stoichiometries, we can use the trends to map out obvious structural
features as a function of rank in order to detect useful descriptors for structure-property relationships. 
Roughly speaking, results Fig.~\ref{fig:patterns} suggest that the energy will typically decrease 
within any given stoichiometry as the number of CC, BN, NN, and BB bonds increases, increases, decreases, decreases, respectively. 
Differences in B and N counts in any ring (a measure of compositional homogeneity per ring),
hardly affects the energy except for the heavily BN doped stoichiometries (11 and 10 BN pairs). 
The degree of clustering (as measured by root mean squared distances) varies wildly with 
little correlation for BB, BN and NN, as long as only few sites have been doped. As the degree of BN-doping increases, the strong stabilising effect of BN bonds subdivides the constitutional isomers into groups of identical BN bond count within which boron clusters are stabilising while nitrogen clusters are destabilising.
Since the BB bond effect is visible for any fixed number of BN bonds, we can conclude that the impact of BN bonds on stability is larger than the one of BB bonds. Following this concept of conditional order, and with the data shown in Fig.~\ref{fig:patterns} and the extended version thereof in the SI, we can identify the following stabilising design patterns in decreasing order of strength: i) add BN pairs, ii) maximize CC bonds, iii) substitute sites shared between rings, iv) maximize BN bonds, v) avoid N substitutions on rings sharing a larger amount of bonds with other rings, and vi) balance BN substitutions in each ring. 
Note how ACE has given us access to a complete ranking without a single quantum chemistry calculation. While the individual rank might not be completely accurate, as shown in Fig.~\ref{fig:ranking}B, the emerging pattern yields relevant and novel structure property trends as a direct consequence of alchemical chirality.

A common problem in materials design consists of identifying global optima. In Fig.~\ref{fig:patterns}, we also
show the least and most stable BN doped derivatives for each stoichiometry, as identified 
by ACE. As more and more carbon sites are BN-doped, the sites interact more strongly and patterns emerge that are in line with the aforementioned summary observations, e.g. the energetically unfavourable nature of nitrogen clusters. 
Thus even if high accuracy is needed eventually, pre-filtering with ACE can dramatically accelerate the identification and discovery of candidate compound lists which are to be treated with higher level quantum  methods subsequently.

\section{Discussion}
We have discussed `alchemical' 4-dimensional chirality, resulting from a reflection in the nuclear charge space manifested by the external potentials in the electronic Hamiltonian. This symmetry relation is exact for the perturbing Hamiltonians of the corresponding enantiomers. The corresponding variational electronic energies are degenerate only up to third order, being reminiscent of parity violation~\cite{QuackPerspectiveParityViolation2011} which also lifts the exact energy degeneracy between spatial enantiomers (albeit less by many orders of magnitude).
Our numerical findings indicate that alchemical chirality-based estimates (ACE) solely from symmetry considerations alone  are sufficiently accurate to enable the exploration of substantial swaths of sub-domains in CCS. From a practical point of view, both experiments and simulations have a resolution limit resulting from method uncertainty beyond which molecules or crystals are indistinguishable. Alchemical chirality offers a new way to find those symmetrically related enantiomers with practically identical energies from which only one needs to be considered in terms of measurements or calculations. In the current state, this could be applied to e.~g. surface adsorption problems like surface catalysis or molecular sensing. Future work on bounds on approximate alchemical enantiomers for relaxed geometries might give access to weak interaction problems as they occur in molecular self-assembly.

Specifically, we have provided novel ACE based bonding rules of chemical bonds and angles. 
Numerical evidence for single bonds of 2nd and 3rd row main-group elements even suggests that DFT 
level of accuracy can be reached with such rules.
CCSD based perturbation theory results for tetrahydrodiazadiborine (and all other possible BN doped benzene derivatives) have served the illustration of alchemical chirality, indicating 
near-degeneracy for the electronic energies of alchemical enantiomers, 
deviating by roughly two orders of magnitude less than a covalent bond. 
Correspondingly, ACE can serve the energy ranking of more complex systems, 
as discussed for the two use-cases of BN doping in naphthalene and in picene. 
Using \mbox{CCSD/cc-pVDZ} as a reference, the comparison to bond-counting, semi-empirical DFT, generalized gradient approximated DFT, and hybrid DFT results for the over 2,000 naphthalene derivatives indicates that the alchemical chirality based ranking outperforms bond counting and common semi-empirical quantum chemistry (PM6), while approaching the performance of semi-empirical DFT (xTB) in terms of fidelity---at negligible computational cost and without empiricism.
For the BN doping of picene, ACE of more than 
400\,M unique derivatives has enabled the establishment of structural trends, as well as the identification of the least and most stable derivatives for each stoichiometry. 
We stress that for all the alchemical chirality derived results presented, not a single quantum chemistry calculation was necessary. 
While  all numerical applications in this study have dealt with alchemical enantiomer pairs of the same stoichiometry (constitutional isomers), they can also differ in stoichiometry. For example, selecting three atoms from an equilateral triangle on the pristine hexagonal Fe (111) surface would account, among many others, also for the alchemical enantiomer pair CoCoCr and MnMnNi.

Overall, our arguments and numerical results indicate that  the concept of alchemical chirality 
represents a novel, fundamental and useful symmetry relation in CCS.
Its power to dissect, group, and rank throughout the CCS of constitutional isomers 
holds great promise to further progress towards the overall goal of virtual computational materials discovery. 
Future work will deal with current limitations, such as the necessity to perform alchemical changes
only in close vicinity in nuclear charge space, to restrict changes to be iso-electronic and neutral, or to rely on scaffold lattices with fixed nuclear positions. 
It would also be interesting to study how alchemical chirality can be exploited using quantum machine learning models.

\section{Materials and Methods}
Previous methodological work tackling chemical compound space from first principles
through variable (`alchemical') nuclear charges included 4-dimensional density functional theory~\cite{WilsonsDFT},
the use of thermodynamic integration~\cite{PolitzerParr}, 
trends among vertical energies~\cite{Mezey1980}, 
entire potential energy surfaces~\cite{Mezey1986}, 
nuclear grand-canonical ensembles~\cite{anatole-prl2005,anatole-jcp2006-2}, 
linear combinations of atomic potentials~\cite{RCD_Yang2006}, and
APDFT~\cite{APDFT}.
Starting in 1996 with stability of solid solutions~\cite{Marzari_prl_1994},
multiple promising applications, based on quantum alchemical changes, 
have been published over recent years, including
thermodynamic integrations~\cite{ArianaAlchemy2006},
mixtures in metal clusters~\cite{AlchemicalDerivativeBinaryMetalCluster_WeigendSchrodtAhlrichs2004,weigend2014extending}, 
reactivity estimates~\cite{CatalystSheppard2010},
water adsorption on BN-doped graphene~\cite{al2017exploring},
BN-doping in fullerenes~\cite{Balawender2018},
protonation energy predictions~\cite{von2020rapidDeprotonation,Munoz2020}
However, apart from nearsightedness studies on chemical transferability~\cite{StijnPNAS2017}, 
general rules, rooted in the quantum mechanical framework of variable composition are mostly still lacking.
Here, we describe `alchemical' chirality, defined by compositional reflection in the nuclear charge mirror plane of some reference system. 
Such a reflection defines alchemical enantiomers as distinct constitutional isomers with electronic energies being identical up to third order. 
Alchemical chirality relates distinct molecules and materials in ways which, to the best of our knowledge, have not yet been discussed before.

Calculating the total potential energy of any compound $U$, most commonly obtained within the Born-Oppenheimer approximation and neglecting relativistic effects, probably represents the most crucial step in any atomistic simulation study. 
It consists of two contributions, the nuclear repulsion which can be evaluated trivially, and the more complex electronic energy $E$ which, within the picture of
density functional theory, sums up the electrons' kinetic contributions as well as their interactions with each other and with the nuclear charges.
As such, $E$ is key, and typically represents the principal goal of most modern electronic structure theory developments, including improved DFT approximations.
But also text-book discussions, such as the virial theorem in physical chemistry, 
deal with the discussion of the chemical bond in terms of the electronic energy.

The difference in electronic energy $E$ between a reference reflection molecule, constituting a maximum in electronic energy, with electron density $\rho_r$ and some ``adjacent'' iso-electronic alchemical enantiomer $i$, $\Delta E = E[\rho_i(\fatr)] - E[\rho_r(\fatr)]$, can be obtained through thermodynamic integration over the coupling constant $0 \le \lambda_i \le 1$ which linearly interpolates the nuclear charges between reflection molecule and alchemical enantiomer. According to Hellmann-Feynman,
$ \Delta E  =  \int_{-\infty}^{+\infty} d\fatr\, \Delta v_{ri}(\fatr)\int_0^1 d\lambda_i \, \rho(\fatr,\lambda_i)$
with $\Delta v_{ri}(\fatr)$ as difference in external potential between reflection molecule and alchemical enantiomer~\cite{APDFT,AtomicAPDFT}.

Approximating this difference by a Taylor series expansion 
($\approx  \sum_{n=1}^{\infty} \frac{1}{n!} \int_{-\infty}^{+\infty} d\fatr\, \Delta v_{ri}(\fatr) \left. \partial^{n-1}_{\lambda_i} \rho(\fatr)\right|_{\lambda = 0}$), 
subtracting the energy of the other alchemical enantiomer $j$ (i.e., $\Delta v_{ri} = - \Delta v_{rj}$), 
and rearranging results in 
\bea
\Delta E^{\rm sym}_{ij} & \approx & \sum_{n=1}^\infty \frac{1}{n!} \int_{-\infty}^{+\infty} d\fatr \,\Delta v_{ri}(\fatr) \left(\frac{\partial^{n-1} \rho_r(\fatr)}{\partial \lambda_i^{n-1}} + \frac{\partial^{n-1} \rho_r(\fatr)}{\partial \lambda_j^{n-1}} \right) \nonumber\\ 
\label{eq:relation}
\eea
which is zero for all orders $n$ as long as
$\partial_{\lambda_i}^{n-1} \rho_r \; = \; -\partial_{\lambda_j}^{n-1} \rho_r$.
Eq.~(\ref{eq:relation}) must be exactly zero up to third order
since i) the zeroth order term $E[\rho_r]$ cancels,
ii) the first order Hellmann-Feynman term $\int d\fatr \Delta v_{ri} \rho_r$ is zero
for highly symmetric systems (as necessary to define an alchemical reflection plane) 
due to $\Delta v$ and $\rho_r$ being odd and even functions~\cite{AlchemicalNormalModes},
and iii) due to the second order term canceling since $\partial_{\lambda} \rho_r = \sum_I \Delta Z_I \partial_{Z_I} \rho_r$ differs for $i$ and $j$ only by the sign of $\Delta Z_I$. Consequently, knowing all the higher order contributions for one alchemical enantiomer also yields all higher order terms for the other enantiomer without any further calculation.
In other words: if a molecule contains a set of disjoint pairs of atoms of the same elements which share
nearly the same chemical environments, it can be seen as the nuclear charge reflection plane, 
or chiral center, of all corresponding alchemical enantiomers.
For example, the benzene molecule is the chiral center for 3 pairs of alchemical enantiomers (for $\Delta Z = \pm 1$, i.e.~BN doping), all discussed in the SI and one highlighted in Figs.~3 and 4.
For iso-electronic charge-neutral mutations, alchemical enantiomers differ from their chiral center only in their nuclear charges
such that the net nuclear charge difference of each atom pair is zero.
We used nauty\cite{McKay2014} for graph enumeration, and the Coulomb Matrix \cite{CM} as implemented in qmlcode\cite{qmlcode2017} as similarity measure. Reference calculations were performed with Molpro\cite{MOLPRO} and MRCC\cite{Nagy2018,mrcc}, in part using basis set extrapolation\cite{Richard2016}.

\begin{acknowledgments}
The authors thank S.~Fias, G.~v.~Hahn, R.~Ramakrishnan, A.~Savin, and J.~ Schrier for discussions.
We acknowledge support by the Swiss National Science foundation (No.~PP00P2\_138932, 407540\_167186 NFP 75 Big Data, 200021\_175747, NCCR MARVEL) and by the European Research Council (ERC-CoG grant QML).
Some calculations were performed at sciCORE (http://scicore.unibas.ch/) scientific computing core facility at University of Basel. The authors declare that they have no competing interests. All data is available in the manuscript, online\cite{sisolving} or the supplementary materials. 

\textbf{Author contributions:}
GvR acquired data and wrote new software used in the work. Both GvR and AvL conceived and planned the project, analyzed and interpreted the results and wrote the manuscript.

\textbf{Supplementary Material}: Details on the naphthalene reference calculations, complete basis set extrapolation, ranking molecules of same stoichiometry, enumeration of molecular graphs, implementation details, site similarity measures, and list of other alchemical bond energy relations.




\end{acknowledgments}

\bibliographystyle{ieeetr}
\bibliography{literatur,main}

\begin{thebibliography}{10}

\bibitem{ShrierNorquist2016ML4failedExperiments}
P.~Raccuglia, K.~C. Elbert, P.~D. Adler, C.~Falk, M.~B. Wenny, A.~Mollo,
  M.~Zeller, S.~A. Friedler, J.~Schrier, and A.~J. Norquist,
  ``Machine-learning-assisted materials discovery using failed experiments,''
  {\em Nature}, vol.~533, no.~7601, pp.~73--76, 2016.

\bibitem{Cronin2020AutonomousRobot}
L.~Porwol, D.~J. Kowalski, A.~Henson, D.-L. Long, N.~L. Bell, and L.~Cronin,
  ``An autonomous chemical robot discovers the rules of inorganic coordination
  chemistry without prior knowledge,'' {\em Angewandte Chemie International
  Edition}, 2020.

\bibitem{Cronin2020RoboticTrends}
P.~S. Gromski, J.~M. Granda, and L.~Cronin, ``Universal chemical synthesis and
  discovery with ‘the chemputer’,'' {\em Trends in Chemistry}, vol.~2,
  no.~1, pp.~4--12, 2020.

\bibitem{Aspuru2020AutonomousExperimentation}
M.~M. Flores-Leonar, L.~M. Mej{\'\i}a-Mendoza, A.~Aguilar-Granda,
  B.~Sanchez-Lengeling, H.~Tribukait, C.~Amador-Bedolla, and A.~Aspuru-Guzik,
  ``Materials acceleration platforms: on the way to autonomous
  experimentation,'' {\em Current Opinion in Green and Sustainable Chemistry},
  p.~100370, 2020.

\bibitem{Aspuru2020selfdrivingDiscoveryThinFilms}
B.~P. MacLeod, F.~G. Parlane, T.~D. Morrissey, F.~H{\"a}se, L.~M. Roch, K.~E.
  Dettelbach, R.~Moreira, L.~P. Yunker, M.~B. Rooney, J.~R. Deeth, {\em
  et~al.}, ``Self-driving laboratory for accelerated discovery of thin-film
  materials,'' {\em Science Advances}, vol.~6, no.~20, p.~eaaz8867, 2020.

\bibitem{pauling1932CovalentBondsPNAS}
L.~Pauling and D.~M. Yost, ``The additivity of the energies of normal covalent
  bonds,'' {\em Proceedings of the National Academy of Sciences of the United
  States of America}, vol.~18, no.~6, p.~414, 1932.

\bibitem{AspuruOpreaRoitbergTetko2020qsar}
E.~N. Muratov, J.~Bajorath, R.~P. Sheridan, I.~V. Tetko, D.~Filimonov,
  V.~Poroikov, T.~I. Oprea, I.~I. Baskin, A.~Varnek, A.~Roitberg, {\em et~al.},
  ``Qsar without borders,'' {\em Chemical Society Reviews}, 2020.

\bibitem{SchneiderReview2010}
G.~Schneider, ``Virtual screening: an endless staircase?,'' {\em Nature
  Reviews}, vol.~9, p.~273, 2010.

\bibitem{mullard2017drug}
A.~Mullard, ``The drug-maker's guide to the galaxy,'' {\em Nature News},
  vol.~549, no.~7673, p.~445, 2017.

\bibitem{Beratan1991}
S.~R. Marder, D.~N. Beratan, and L.-T. Cheng, ``Approaches for optimizing the
  first electronic hyperpolarizability of conjugated organic molecules,'' {\em
  Science}, vol.~252, pp.~103--106, 1991.

\bibitem{ceder1998predicting}
G.~Ceder, ``Predicting properties from scratch,'' {\em Science}, vol.~280,
  no.~5366, pp.~1099--1100, 1998.

\bibitem{ZungerPRL1998}
C.~Wolverton and A.~Zunger, ``First-principles prediction of vacancy
  order-disorder and intercalation battery voltages in {LixCoO2},'' {\em Phys.
  Rev. Lett.}, vol.~81, p.~606, 1998.

\bibitem{ZungerNature1999}
A.~Franceschetti and A.~Zunger, ``The inverse band-structure problem of finding
  an atomic configuration with given electronic properties,'' {\em Nature},
  vol.~402, p.~60, 1999.

\bibitem{alberi20182019}
K.~Alberi, M.~B. Nardelli, A.~Zakutayev, L.~Mitas, S.~Curtarolo, A.~Jain,
  M.~Fornari, N.~Marzari, I.~Takeuchi, M.~L. Green, {\em et~al.}, ``The 2019
  materials by design roadmap,'' {\em Journal of Physics D: Applied Physics},
  vol.~52, no.~1, p.~013001, 2018.

\bibitem{HK}
P.~Hohenberg and W.~Kohn, ``Inhomogeneous electron gas,'' {\em Phys. Rev.},
  vol.~136, p.~B864, 1964.

\bibitem{PBE}
J.~P. Perdew, K.~Burke, and M.~Ernzerhof, ``Generalized gradient approximation
  made simple,'' {\em Phys. Rev. Lett.}, vol.~77, p.~3865, 1996.

\bibitem{ComputationalMaterialsDesign_MRS2006}
J.~Hafner, C.~Wolverton, G.~Ceder, and G.~Editors, ``Toward computational
  materials design: {The} impact of density functional theory on materials
  research,'' {\em MRS Bulletin}, vol.~31, p.~659, 2006.

\bibitem{BurkePerspectives_2012jcp}
K.~Burke, ``Perspective on density functional theory,'' {\em J. Chem. Phys.},
  vol.~136, no.~15, p.~150901, 2012.

\bibitem{marzari2016materials}
N.~Marzari, ``Materials modelling: The frontiers and the challenges,'' {\em
  Nature materials}, vol.~15, no.~4, p.~381, 2016.

\bibitem{QMLessayAnatole}
O.~A. von Lilienfeld, ``Quantum machine learning in chemical compound space,''
  {\em Angewandte Chemie International Edition}, vol.~57, p.~4164, 2018.
\newblock http://dx.doi.org/10.1002/anie.201709686.

\bibitem{Anatole2020NatureReview}
O.~A. von Lilienfeld, K.-R. M{\"u}ller, and A.~Tkatchenko, ``Exploring chemical
  compound space with quantum-based machine learning,'' {\em Nature Reviews
  Chemistry}, pp.~1--12, 2020.

\bibitem{MachineLearningMeetsQuantumPhysics2020book}
K.~Sch{\"u}tt, S.~Chmiela, O.~von Lilienfeld, A.~Tkatchenko, K.~Tsuda, and
  K.~M{\"u}ller, {\em Machine Learning Meets Quantum Physics}.
\newblock Lecture Notes in Physics, Springer International Publishing, 2020.

\bibitem{RampiMLQMMM}
V.~Botu and R.~Ramprasad, ``Adaptive machine learning framework to accelerate
  ab initio molecular dynamics,'' {\em Int. J. Quantum Chem.}, vol.~115,
  no.~16, pp.~1074--1083, 2015.

\bibitem{CovariantKernelsSandro2016}
A.~Glielmo, P.~Sollich, and A.~De~Vita, ``Accurate interatomic force fields via
  machine learning with covariant kernels,'' {\em Physical Review B}, vol.~95,
  no.~21, p.~214302, 2017.

\bibitem{chmiela2017machine}
S.~Chmiela, A.~Tkatchenko, H.~E. Sauceda, I.~Poltavsky, K.~T. Sch{\"u}tt, and
  K.-R. M{\"u}ller, ``Machine learning of accurate energy-conserving molecular
  force fields,'' {\em Science Advances}, vol.~3, no.~5, p.~e1603015, 2017.

\bibitem{grisafi2018symmetry}
A.~Grisafi, D.~M. Wilkins, G.~Cs\'anyi, and M.~Ceriotti, ``Symmetry-adapted
  machine learning for tensorial properties of atomistic systems,'' {\em Phys.
  Rev. Lett.}, vol.~120, p.~036002, Jan 2018.

\bibitem{chmiela2019sgdml}
S.~Chmiela, H.~E. Sauceda, I.~Poltavsky, K.-R. M{\"u}ller, and A.~Tkatchenko,
  ``sgdml: Constructing accurate and data efficient molecular force fields
  using machine learning,'' {\em Computer Physics Communications}, 2019.

\bibitem{OQML}
A.~S. Christensen, F.~A. Faber, and O.~A. von Lilienfeld, ``Operators in
  quantum machine learning: Response properties in chemical space,'' {\em The
  Journal of Chemical physics}, vol.~150, no.~6, p.~064105, 2019.

\bibitem{Amons}
B.~Huang and O.~A. von Lilienfeld, ``Quantum machine learning using
  atom-in-molecule-based fragments selected on the fly,'' {\em Nature
  Chemistry}, 2020.
\newblock in press.

\bibitem{anatole-ijqc2013}
O.~A. von Lilienfeld, ``First principles view on chemical compound space:
  Gaining rigorous atomistic control of molecular properties,'' {\em
  International Journal of Quantum Chemistry}, vol.~113, no.~12,
  pp.~1676--1689, 2013.

\bibitem{APDFT}
G.~F. von Rudorff and O.~A. von Lilienfeld, ``Alchemical perturbation density
  functional theory,'' {\em Phys. Rev. Research}, vol.~2, p.~023220, 2020.

\bibitem{Elguiero2005BDEdiatomicMaingroup}
O.~M{\'o}, M.~Y{\'a}{\~n}ez, M.~Eckert-Maksi{\'c}, Z.~B. Maksi{\'c},
  I.~Alkorta, and J.~Elguero, ``Periodic trends in bond dissociation energies.
  a theoretical study,'' {\em The Journal of Physical Chemistry A}, vol.~109,
  no.~19, pp.~4359--4365, 2005.

\bibitem{ChemistsGuidetoDFT}
W.~Koch and M.~C. Holthausen, {\em A Chemist's Guide to Density Functional
  Theory}.
\newblock Wiley-VCH, 2002.

\bibitem{Ryczko2020}
K.~Ryczko, P.~Darancet, and I.~Tamblyn, ``Inverse design of a graphene-based
  quantum transducer via neuroevolution,'' {\em The Journal of Physical
  Chemistry C}, vol.~124, pp.~26117--26123, nov 2020.

\bibitem{Keith2017alchemicalCatalysis}
K.~Saravanan, J.~R. Kitchin, O.~A. von Lilienfeld, and J.~A. Keith,
  ``Alchemical predictions for computational catalysis: Potential and
  limitations,'' {\em The Journal of Physical Chemistry Letters}, vol.~8,
  no.~20, pp.~5002--5007, 2017.

\bibitem{Keith2018benchmarkingalchemy}
C.~D. Griego, K.~Saravanan, and J.~A. Keith, ``Benchmarking computational
  alchemy for carbide, nitride, and oxide catalysts,'' {\em Advanced Theory and
  Simulations}, p.~1800142, 2018.

\bibitem{Keith2020acceleratedCatalystAlchemy}
C.~D. Griego, J.~R. Kitchin, and J.~A. Keith, ``Acceleration of catalyst
  discovery with easy, fast, and reproducible computational alchemy,'' {\em
  International Journal of Quantum Chemistry}, p.~e26380, 2020.

\bibitem{PorousBNnanofibers2020graphene}
H.~Wang, J.~Liang, Y.~Wu, T.~Kang, D.~Shen, Z.~Tong, R.~Yang, Y.~Jiang, D.~Wu,
  X.~Li, {\em et~al.}, ``Porous bn nanofibers enable long-cycling life sodium
  metal batteries,'' {\em Small}, p.~2002671, 2020.

\bibitem{Marzari20202highthroughput2Dmaterials}
N.~Mounet, M.~Gibertini, P.~Schwaller, D.~Campi, A.~Merkys, A.~Marrazzo,
  T.~Sohier, I.~E. Castelli, A.~Cepellotti, G.~Pizzi, {\em et~al.},
  ``Two-dimensional materials from high-throughput computational exfoliation of
  experimentally known compounds,'' {\em Nature nanotechnology}, vol.~13,
  no.~3, pp.~246--252, 2018.

\bibitem{Raghu2019polycyclicBNdoped}
S.~Chakraborty, P.~Kayastha, and R.~Ramakrishnan, ``The chemical space of b,
  n-substituted polycyclic aromatic hydrocarbons: Combinatorial enumeration and
  high-throughput first-principles modeling,'' {\em The Journal of chemical
  physics}, vol.~150, no.~11, p.~114106, 2019.

\bibitem{Samuel-CHIMIA2014}
K.~Chang and O.~A. von Lilienfeld, ``Quantum mechanical treatment of variable
  molecular composition: From'alchemical'changes of state functions to rational
  compound design,'' {\em CHIMIA International Journal for Chemistry}, vol.~68,
  no.~9, pp.~602--608, 2014.

\bibitem{Samuel-JCP2016}
K.~Y.~S. Chang, S.~Fias, R.~Ramakrishnan, and O.~A. von Lilienfeld, ``Fast and
  accurate predictions of covalent bonds in chemical space,'' {\em J. Chem.
  Phys.}, vol.~144, no.~17, p.~174110, 2016.

\bibitem{MoritzBaben-JCP2016}
M.~to~Baben, J.~O. Achenbach, and O.~A. von Lilienfeld, ``Guiding ab initio
  calculations by alchemical derivatives,'' {\em J. Chem. Phys.}, vol.~144,
  p.~104103, 2016.

\bibitem{AlchemyAlisa_2016}
A.~Solovyeva and O.~A. von Lilienfeld, ``Alchemical screening of ionic
  crystals,'' {\em Phys. Chem. Chem. Phys.}, vol.~18, pp.~31078--31091, 2016.

\bibitem{Samuel2018bandgaps}
K.~S. Chang and O.~A. von Lilienfeld, ``{Al}$_x${Ga}$_{1-x}${As} crystals with
  direct 2 {eV} band gaps from computational alchemy,'' {\em Physical Review
  Materials}, vol.~2, no.~7, p.~073802, 2018.

\bibitem{AlchemicalNormalModes}
S.~Fias, K.~S. Chang, and O.~A. von Lilienfeld, ``Alchemical normal modes unify
  chemical space,'' {\em The journal of physical chemistry letters}, vol.~10,
  pp.~30--39, 2018.

\bibitem{ReviewCatalystNorskov}
J.~K. N{\o}rskov, T.~Bligaard, J.~Rossmeisl, and C.~H. Christensen, ``Towards
  the computational design of solid catalysts,'' {\em Nature Chemistry},
  vol.~1, p.~37, 2009.

\bibitem{MaterialsProject}
S.~P. Ong, A.~Jain, G.~Hautier, M.~Kocher, S.~Cholia, D.~Gunter, D.~Bailey,
  D.~Skinner, K.~A. Persson, and G.~Ceder, ``{The Materials Project},'' 2011.
\newblock http://materialsproject.org/.

\bibitem{Mezey1998}
P.~G. Mezey {\em Journal of Mathematical Chemistry}, vol.~23, no.~1/2,
  pp.~65--84, 1998.

\bibitem{Mezey1980}
P.~G. Mezey, ``Electronic energy inequalities for isoelectronic molecular
  systems,'' {\em Theoretica Chimica Acta}, vol.~59, no.~4, pp.~321--332, 1980.

\bibitem{Daza2005}
E.~E. Daza and A.~Bernal, ``Energy bounds for isoelectronic molecular sets and
  the implicated order,'' {\em J. Math. Chem.}, vol.~38, pp.~247--263, aug
  2005.

\bibitem{Chakraborty2019}
S.~Chakraborty, P.~Kayastha, and R.~Ramakrishnan, ``The chemical space of b,
  n-substituted polycyclic aromatic hydrocarbons: Combinatorial enumeration and
  high-throughput first-principles modeling,'' {\em The Journal of Chemical
  Physics}, vol.~150, p.~114106, mar 2019.

\bibitem{PBE0}
J.~P. Perdew, M.~Ernzerhof, and K.~Burke {\em J. Chem. Phys.}, vol.~105,
  p.~9982, 1996.

\bibitem{PBE01}
M.~Ernzerhof and G.~E. Scuseria {\em J. Chem. Phys.}, vol.~110, p.~5029, 1999.

\bibitem{PBE02}
C.~Adamo and V.~Barone, ``Toward reliable density functional methods without
  adjustable parameters: The pbe0 model,'' {\em J. Chem. Phys.}, vol.~110,
  no.~13, pp.~6158--6170, 1999.

\bibitem{Grimme2019GFn2xTB}
C.~Bannwarth, S.~Ehlert, and S.~Grimme, ``Gfn2-xtb—an accurate and broadly
  parametrized self-consistent tight-binding quantum chemical method with
  multipole electrostatics and density-dependent dispersion contributions,''
  {\em Journal of chemical theory and computation}, vol.~15, no.~3,
  pp.~1652--1671, 2019.

\bibitem{PM6}
J.~J.~P. Stewart, ``{Optimization of parameters for semiempirical methods V:
  Modification of NDDO approximations and application to 70 elements},'' {\em
  Journal of Molecular Modeling}, vol.~13, no.~12, pp.~1173--1213, 2007.

\bibitem{UFFRappeGoddard1992}
A.~K. Rapp\'e, C.~J. Casewit, K.~S. Colwell, W.~A.~G. III, and W.~M. Skid,
  ``{UFF}, a full periodic table force field for molecular mechanics and
  molecular dynamics simulations,'' {\em J. Am. Chem. Soc.}, vol.~114,
  p.~10024, 1992.

\bibitem{QuackPerspectiveParityViolation2011}
M.~Quack, ``Frontiers in spectroscopy,'' {\em Faraday Discussions}, vol.~150,
  pp.~533--565, 2011.

\bibitem{WilsonsDFT}
{E. B. Wilson, Jr.}, ``Four dimensional electron density function,'' {\em J.
  Chem. Phys.}, vol.~36, p.~2232, 1962.

\bibitem{PolitzerParr}
P.~Politzer and R.~G. Parr, ``Some new energy formulas for atoms and
  molecules,'' {\em J. Chem. Phys.}, vol.~61, p.~4258, 1974.

\bibitem{Mezey1986}
P.~G. Mezey, ``New global constraints on electronic energy hypersurfaces,''
  {\em International Journal of Quantum Chemistry}, vol.~29, pp.~85--99, jan
  1986.

\bibitem{anatole-prl2005}
O.~A. von Lilienfeld, R.~Lins, and U.~Rothlisberger, ``Variational particle
  number approach for rational compound design,'' {\em Phys. Rev. Lett.},
  vol.~95, p.~153002, 2005.

\bibitem{anatole-jcp2006-2}
O.~A. von Lilienfeld and M.~E. Tuckerman, ``Molecular grand-canonical ensemble
  density functional theory and exploration of chemical space,'' {\em J. Chem.
  Phys.}, vol.~125, p.~154104, 2006.

\bibitem{RCD_Yang2006}
M.~Wang, X.~Hu, D.~N. Beratan, and W.~Yang, ``Designing molecules by optimizing
  potentials,'' {\em J. Am. Chem. Soc.}, vol.~128, p.~3228, 2006.

\bibitem{Marzari_prl_1994}
N.~Marzari, S.~de~Gironcoli, and S.~Baroni, ``Structure and phase stability of
  {G}a$_x${I}n$_{1-x}${P} solid solutions from computational alchemy,'' {\em
  Phys. Rev. Lett.}, vol.~72, p.~4001, 1994.

\bibitem{ArianaAlchemy2006}
A.~Beste, R.~J. Harrison, and T.~Yanai, ``Direct computation of general
  chemical energy differences: Application to ionization potentials,
  excitation, and bond energies,'' {\em J. Phys. Chem.}, vol.~125, p.~074101,
  2006.

\bibitem{AlchemicalDerivativeBinaryMetalCluster_WeigendSchrodtAhlrichs2004}
F.~Weigend, C.~Schrodt, and R.~Ahlrichs, ``Atom distributions in binary atom
  clusters: {A} perturbational approach and its validation in a case study,''
  {\em J. Chem. Phys.}, vol.~121, p.~10380, 2004.

\bibitem{weigend2014extending}
F.~Weigend, ``Extending dft-based genetic algorithms by atom-to-place
  re-assignment via perturbation theory: a systematic and unbiased approach to
  structures of mixed-metallic clusters,'' {\em The Journal of chemical
  physics}, vol.~141, no.~13, p.~134103, 2014.

\bibitem{CatalystSheppard2010}
D.~Sheppard, G.~Henkelman, and O.~A. von Lilienfeld, ``Alchemical derivatives
  of reaction energetics,'' {\em J. Chem. Phys.}, vol.~133, p.~084104, 2010.

\bibitem{al2017exploring}
Y.~S. Al-Hamdani, A.~Michaelides, and O.~A. von Lilienfeld, ``Exploring
  dissociative water adsorption on isoelectronically bn doped graphene using
  alchemical derivatives,'' {\em The Journal of chemical physics}, vol.~147,
  no.~16, p.~164113, 2017.

\bibitem{Balawender2018}
R.~Balawender, M.~Lesiuk, F.~D. Proft, and P.~Geerlings, ``Exploring chemical
  space with alchemical derivatives: {BN}-simultaneous substitution patterns in
  c60,'' {\em J. Chem. Theory Comput.}, vol.~14, pp.~1154--1168, jan 2018.

\bibitem{von2020rapidDeprotonation}
G.~F. von Rudorff and O.~A. von Lilienfeld, ``Rapid and accurate molecular
  deprotonation energies from quantum alchemy,'' {\em Physical Chemistry
  Chemical Physics}, vol.~22, pp.~10519--10525, 2020.

\bibitem{Munoz2020}
M.~Mu{\~{n}}oz, A.~Robles-Navarro, P.~Fuentealba, and C.~C{\'{a}}rdenas,
  ``Predicting deprotonation sites using alchemical derivatives,'' {\em The
  Journal of Physical Chemistry A}, vol.~124, pp.~3754--3760, apr 2020.

\bibitem{StijnPNAS2017}
S.~Fias, F.~Heidar-Zadeh, P.~Geerlings, and P.~W. Ayers, ``Chemical
  transferability of functional groups follows from the nearsightedness of
  electronic matter,'' {\em Proceedings of the National Academy of Sciences},
  vol.~114, no.~44, pp.~11633--11638, 2017.

\bibitem{AtomicAPDFT}
G.~F. von Rudorff and O.~A. von Lilienfeld, ``Atoms in molecules from
  alchemical perturbation density functional theory,'' {\em The Journal of
  Physical Chemistry B}, vol.~123, pp.~10073--10082, 2019.

\bibitem{McKay2014}
B.~D. McKay and A.~Piperno, ``Practical graph isomorphism, {II},'' {\em Journal
  of Symbolic Computation}, vol.~60, pp.~94--112, jan 2014.

\bibitem{CM}
M.~Rupp, A.~Tkatchenko, K.-R. M\"uller, and O.~A. von Lilienfeld, ``Fast and
  accurate modeling of molecular atomization energies with machine learning,''
  {\em Phys. Rev. Lett.}, vol.~108, p.~058301, 2012.

\bibitem{qmlcode2017}
A.~S. Christensen, F.~A. Faber, B.~Huang, L.~A. Bratholm, A.~Tkatchenko, K.-R.
  M\"uller, and O.~A. von Lilienfeld, ``Qml: A python toolkit for quantum
  machine learning,'' 2017.

\bibitem{MOLPRO}
H.-J. Werner, P.~J. Knowles, G.~Knizia, F.~R. Manby, M.~{Sch\"{u}tz},
  P.~Celani, W.~Gy\"orffy, D.~Kats, T.~Korona, R.~Lindh, A.~Mitrushenkov,
  G.~Rauhut, K.~R. Shamasundar, T.~B. Adler, R.~D. Amos, S.~J. Bennie,
  A.~Bernhardsson, A.~Berning, D.~L. Cooper, M.~J.~O. Deegan, A.~J. Dobbyn,
  F.~Eckert, E.~Goll, C.~Hampel, A.~Hesselmann, G.~Hetzer, T.~Hrenar,
  G.~Jansen, C.~K\"oppl, S.~J.~R. Lee, Y.~Liu, A.~W. Lloyd, Q.~Ma, R.~A. Mata,
  A.~J. May, S.~J. McNicholas, W.~Meyer, T.~F. {Miller III}, M.~E. Mura,
  A.~Nicklass, D.~P. O'Neill, P.~Palmieri, D.~Peng, K.~Pfl\"uger, R.~Pitzer,
  M.~Reiher, T.~Shiozaki, H.~Stoll, A.~J. Stone, R.~Tarroni, T.~Thorsteinsson,
  M.~Wang, and M.~Welborn, ``Molpro, version 2018.1, a package of ab initio
  programs,'' 2018.

\bibitem{Nagy2018}
P.~R. Nagy, G.~Samu, and M.~K{\'{a}}llay, ``Optimization of the linear-scaling
  local natural orbital {CCSD}(t) method: Improved algorithm and benchmark
  applications,'' {\em Journal of Chemical Theory and Computation}, vol.~14,
  pp.~4193--4215, jul 2018.

\bibitem{mrcc}
M.~K\'allay, P.~R. Nagy, Z.~Rolik, D.~Mester, G.~Samu, J.~Csontos, J.~Cs\'oka,
  B.~P. Szab\'o, L.~Gyevi-Nagy, I.~Ladj\'anszki, L.~Szegedy, B.~Lad\'oczki,
  K.~Petrov, M.~Farkas, P.~D. Mezei, and B.~H\'egely, ``{Mrcc, a quantum
  chemical program suite, www.mrcc.hu},'' 2019.

\bibitem{Richard2016}
R.~M. Richard, M.~S. Marshall, O.~Dolgounitcheva, J.~V. Ortiz, J.-L.
  Br{\'{e}}das, N.~Marom, and C.~D. Sherrill, ``Accurate ionization potentials
  and electron affinities of acceptor molecules i. reference data at the
  {CCSD}(t) complete basis set limit,'' {\em Journal of Chemical Theory and
  Computation}, vol.~12, pp.~595--604, jan 2016.

\bibitem{sisolving}
G.~F. von Rudorff and O.~A. von Lilienfeld, ``Solving the inverse materials
  design problem with alchemical chirality,'' {\em 10.5281/zenodo.3994178}, Aug
  2020.

\end{thebibliography}


\end{document}